\documentclass[twocolumn,showpacs,superscriptaddress,preprintnumbers,amsmath,amssymb]{revtex4}

\usepackage{graphicx}
\usepackage{dcolumn}
\usepackage{amsmath}
\usepackage{amssymb}

\def\vq{{\bf q}}

\def\vk{{\bf k}}

\def\vr{{\bf r}}
\def\vS{{\bf S}}

\def\bra{\langle}
\def\ket{\rangle}

\newcommand{\fig}[1]{Fig.~\ref{#1}}
\newcommand{\be}{\begin{equation}}
\newcommand{\ee}{\end{equation}}
\newcommand{\bea}{\begin{eqnarray}}

\newcommand{\eea}{\end{eqnarray}}
\newcommand{\bean}{\begin{eqnarray*}}
\newcommand{\eean}{\end{eqnarray*}}
\newcommand{\bfi}{\begin{figure}}
\newcommand{\efi}{\end{figure}}
\newcommand{\bc}{\begin{center}}
\newcommand{\ec}{\end{center}}
\newcommand{\ba}{\begin{array}}
\newcommand{\ea}{\end{array}}

\begin{document}


\title{Orientational symmetry-breaking correlations  
in square lattice $t$-$J$ model}

\author{Akiharu Miyanaga} 
\affiliation{Spread society of practical algorithm and software,  
3-2-16 Chuo, Ebina, Kanagawa 243-0432, Japan}
\author{Hiroyuki Yamase} 
\affiliation{Max-Planck-Institute for Solid State Research, 
Heisenbergstrasse 1, D-70569 Stuttgart, Germany}

\date{\today}

\begin{abstract}
We study a tendency to orientational symmetry breaking 
of the square lattice $t$-$J$ model by applying 
the exact diagonalization technique.  
We introduce a small external anisotropy to $t$ and $J$, and 
calculate the induced anisotropy of hole-hole correlations as a response.
It is found that the response is 
strongly enhanced for particular band parameters, meaning that 
the system has a strong tendency to 
orientational symmetry breaking. 
The analysis of the momentum distribution function indicates that the  
correlation develops when the Fermi surface is close to 
$(\pi,\,0)$ and $(0,\,\pi)$. 
Some properties of high-$T_{c}$ cuprates previously attributed to 
stripes can also be understood from orientational symmetry breaking alone. 
\end{abstract}

\maketitle

High-$T_{c}$ cuprate superconductors are realized by a carrier doping 
into Mott insulators, 
where strong onsite Coulomb repulsion prohibits double occupation 
and one electron resides at every sites. 
Such strong electron-electron correlations are 
widely recognized to be crucial to the physics of high-$T_{c}$ cuprates.

Minimal models for high-$T_{c}$ cuprates are believed to be 
the two-dimensional (2D) $t$-$J$ and Hubbard models on a square lattice. 
Cuprate superconductors have $d$-wave gap symmetry,\cite{harlingen-tsuei} 
which was successfully reproduced in these models. 
The models, however, exhibit several ordering tendencies competing with 
$d$-wave superconductivity: antiferromagnetism,\cite{inui88,giamarchi91}   
spin-charge stripes,\cite{white98} 
staggered flux,\cite{wen96} and  
$d$-density wave.\cite{zeyher99} 
These possible competing orders were recognized as playing 
crucial roles in high-$T_{c}$ cuprates and discussed 
in relation to pseudogap phenomena,\cite{timusk99} 
vortex structures,\cite{demler04,lee01} and 
charge ordering.\cite{kivelson03} 

Recently, through microscopic analysises of the $t$-$J$\cite{yamase00} 
and Hubbard\cite{metzner00,valenzuela01,wegner0203,neumayr03} models 
another competing 
order was proposed, the $d$-wave Fermi surface deformation ($d$FSD): 
the Fermi surface (FS) expands along the $k_{x}$ direction 
and shrinks along the $k_{y}$ direction or vice versa. 
This order is generated 
by forward scattering of electrons close to the FS near 
$(\pi,\,0)$ and $(0,\,\pi)$, and hence should be distinguished from all 
the competing orders mentioned above, which 
require large momentum transfer with $\vq \approx (\pi,\,\pi)$. 

Some authors termed the spontaneous $d$FSD as Pomeranchuk instability, 
referring to a stability criterion for normal Fermi liquids by 
Pomeranchuk.\cite{pomeranchuk58}   
Note, however, that the $d$FSD turned out to occur usually through a 
first order transition at low temperature, namely 
without breaking Pomeranchuk's stability 
condition.\cite{kee0304,yamase05a} 

The $d$FSD competes with other instabilities. 
In particular, it can be overwhelmed 
by $d$-wave singlet pairing.\cite{yamase00,honerkamp02}  
However, it was pointed out that even if the spontaneous $d$FSD was 
prohibited, the correlations survived\cite{yamase04b} and 
made a system sensitive to an external anisotropy, 
which led to a noticeable $d$FSD.\cite{yamase00,metzner03} 
This scenario was invoked for high-$T_{c}$ cuprates
through the slave-boson mean-field analysis of the 
$t$-$J$ model.\cite{yamase00} 
While the slave-boson theory treats strong correlation 
effects on average, it still remains a fundamental question 
whether the $d$FSD will be relevant to the strongly correlated regime 
and thus to the physics of high-$T_{c}$ cuprates. 
Since the $d$FSD changes the shape of the FS, 
possibilities of the $d$FSD are crucial to the understanding 
of low energy properties of cuprates.

In this Letter, we study the $d$FSD by applying 
the exact diagonalization technique to the 2D $t$-$J$ model at 
zero temperature and 
treat strong correlation effects rigorously. 
We take a $4\times 4$ cluster with a periodic boundary condition 
with two holes. 
Since the $d$FSD is associated with 
orientational symmetry breaking  of a square lattice, 
we focus on a tendency to the orientational symmetry breaking. 
We impose a small 
$xy$-spatial anisotropy in the original $t$-$J$ model and 
calculate the induced anisotropy of 
hole-hole correlations as a response. 
It is found that the response shows a sharp peak 
as a function of band parameters, which means that the system 
has a strong tendency to orientational symmetry breaking 
for particular band parameters.  
The analysis of the momentum distribution function indicates that 
this correlation develops when the FS is close to 
$(\pi,\,0)$ and $(0,\,\pi)$. 
The concept of the $d$FSD, 
originally proposed in the slave-boson mean-field analysis of the 
$t$-$J$ model\cite{yamase00} and 
a renormalization group analysis of the Hubbard model,\cite{metzner00} 
is thus relevant even to the strongly correlated regime and 
in this sense is a generic feature of square lattice interacting 
electron systems. Possibilities of the $d$FSD are discussed for 
high-$T_{c}$ cuprates.

We take the 2D $t$-$J$ model 
\be
 H = -  \sum_{i,\,\tau,\, \sigma} t^{(l)}_{\tau} 
 \tilde{c}_{i\,\sigma}^{\dagger}\tilde{c}_{i+\tau\,\sigma}+
   \sum_{\langle i,i+\tau \rangle}J_{\tau} \left(
\vS_{i} \cdot \vS_{i+\tau}-\frac{1}{4}\tilde{n}_{i}\tilde{n}_{i+\tau} 
\right)  \label{tJ} 
\ee  
defined in the Fock space with no doubly occupied sites. 
The $\tilde{c}_{i\,\sigma}$ ($\vS_{i}$) is an electron (a spin) 
operator, and 
$\tilde{n}_{i}=\sum_{\sigma}\tilde{c}_{i\sigma}^{\dagger}\tilde{c}_{i\sigma}$ is the number operator of electrons at site $i$.  
The $t^{(l)}_{\tau}$ is the $l$th ($l \leq 3$) 
neighbor hopping integral to the direction 
$\boldsymbol\tau={\bf r}_{j}-{\bf r}_{i}$, 
and we denote $t^{(1)}_{\tau}=t_{\tau}$, $t^{(2)}_{\tau}=t'$, 
and  $t^{(3)}_{\tau}=t''$; the latter two are  assumed isotropic.  
The $J_{\tau}(>0)$ is the superexchange coupling 
between nearest-neighbor sites. 
We introduce an external anisotropy to $t_{\tau}$ and $J_{\tau}$, 
\bea
& & t_{x}=(1+\alpha) t\, , \; t_{y}=(1-\alpha) t\,,\\
& & J_{x}=(1+\gamma) J\, , \; J_{y}=(1-\gamma) J\,,
\eea
where we assume $\gamma=\frac{2\alpha}{1-\alpha}$ so that 
the anisotropy between $J_{x}$ and $J_{y}$ is twice as large as 
that between $t_{x}$ and $t_{y}$. 
As a response to this anisotropy, 
we consider an anisotropy of hole-hole correlation 
\be
\Delta\! C(\vr) = C({\bf r})-C(\bar{\bf r}) \,,
\ee
where ${\bf r}=(x,y)$ and $\bar{\bf r}=(y,x)$, and 
\be 
C({\bf r}) = \frac{1}{(N\delta)^{2}}
\sum_{i}\bra(1-\tilde{n}_{i})(1-\tilde{n}_{i+r})\ket\, 
\ee
is a hole-hole correlation function;
$N$ is the total number of lattice sites; $\delta(>0)$ is 
hole density; the factor $\frac{1}{(N\delta)^{2}}$ is taken to normalize 
$\sum_{\vr}C(\vr)$ to be unity. 
Note that $\Delta\! C(\vr)$ vanishes as long as the system has 
orientational symmetry of the lattice.

\begin{figure}[t]
\centerline{\includegraphics[width=0.45\textwidth]{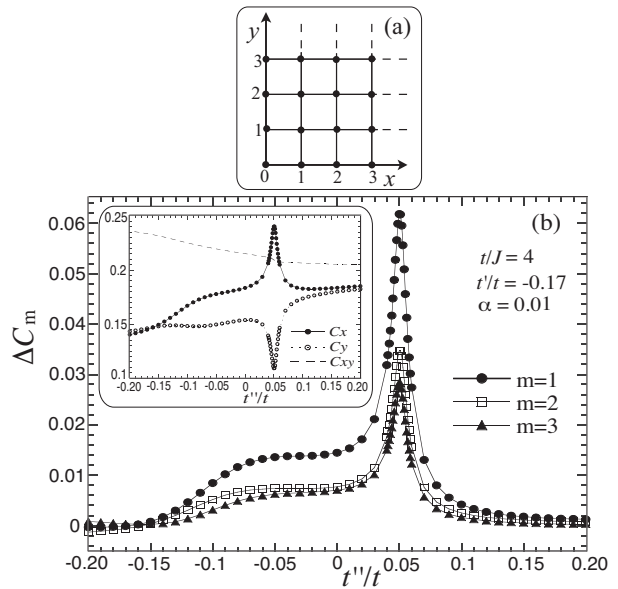}}
%
\caption{(a) The coordinates of the $4\times 4$ cluster. 
(b) $t''$ dependence of the anisotropy of hole-hole correlation  
$\Delta\!C_{m}$ for $\alpha=0.01$, $t/J=4$, and $t'/t=-0.17$. 
The inset shows 
hole-hole correlations along the $x(y)$ direction $C_{x(y)}$  
and a diagonal direction $C_{xy}$. 
}
\label{Dhh-t2}
\end{figure}

To treat strong correlation effects rigorously, we take  
a $4\times 4$ cluster with a periodic boundary condition 
and diagonalize the Hamiltonian 
using the Lanczos algorithm at zero temperature; 
the coordinates of the cluster are defined in \fig{Dhh-t2}(a). 
Although the cluster size is rather small for usual exact 
diagonalization studies, the $4\times 4$ cluster 
has square lattice symmetry and enables us to  
treat $\vk=(\pi,\,0)$ and $(0,\,\pi)$. These aspects are crucial in the 
present study. 
We introduce two holes ($\delta=1/8$) and 
assume that the total momentum of the wave function is zero. 
For $\alpha=0$, the system has orientational symmetry 
and hence $\Delta\! C(\vr)=0$. 
When a small $\alpha$ is introduced, $\Delta\! C(\vr)$ can become finite. 
It has four inequivalent points: 
(i) $\vr=(2,0)$, (ii) $\vr=(1,0)$, $(3,0)$, 
(iii) $\vr=(2,1)$, $(2,3)$, and (iv) $\vr=(1,1)$, $(2,2)$, $(3,3)$,
$(3,1)$.\cite{miscr} 
In the latter case $\Delta\! C (\vr)=0$. We refer to  
the former three cases as $\Delta\! C_{m}$ 
with $m=1,2$, and $3$, respectively. 
Setting $\alpha=0.01$,  
we calculate $\Delta\! C_{m}$  
as a function of $t''$ for $t/J=4$ and $t'/t=-0.17$.

In \fig{Dhh-t2}(b), we see that $\Delta\! C_{m}$ is nearly zero 
for $|t''|\gtrsim 0.15t$, 
that is the hole-hole correlation is almost isotropic even in the 
presence of the external anisotropy ($\alpha \neq 0$). 
For $|t''| \lesssim 0.1t$, $\Delta\! C_{m}$ becomes noticeable and is 
strongly enhanced near  $t''/t= 0.05$ to form a sharp peak. 
This peak is seen in all $\Delta\! C_{m}$, indicating that 
the system has a strong tendency to 
the  orientational symmetry breaking for particular band parameters.

In the inset of \fig{Dhh-t2}(b), 
we show the following quantity as a function of 
$t''$ to see a degree of the orientational symmetry breaking: 
$C_{x}=\sum_{x}C(\vr)$, where the summation is over 
$\vr=(1,0), (2,0),(3,0)$, similarly 
$C_{y}=\sum_{y}C(\vr)$ over $\vr=(0,1), (0,2),(0,3)$, and 
$C_{xy}=\sum_{xy}C(\vr)$ over $\vr=(1,1), (2,2),(3,3)$. 
The $C_{x}$ ($C_{y}$) and $C_{xy}$ describe hole-hole correlation 
along the  horizontal (vertical) direction and a diagonal direction, 
respectively. Around $t''/t=0.05$ 
the value of $C_{x}$ is about twice as 
large as that of $C_{y}$. The breaking of orientational symmetry 
at $t''/t\approx 0.05$ is thus very large. 
The hole-hole correlation, however, does not 
become one-dimensional like. 
The sum rule, $\sum_{\vr}C(\vr)=1$, suggests that 
the value of $C_{xy}\approx 0.2$ is appreciable since 
the mean value of $C(\vr)$ is $\frac{1}{15}$ and 
that of $C_{xy}$ becomes $\frac{3}{15}$. 
Furthermore we found that 
$C_{xy}$ was almost the same as that for $\alpha=0$, 
that is the correlation along the diagonal direction still remained. 
Moving away from $t''/t\approx 0.05$ the difference between 
$C_{x}$ and $C_{y}$ gets smaller as expected.

\begin{figure}[t]
\centerline{\includegraphics[width=0.45\textwidth]{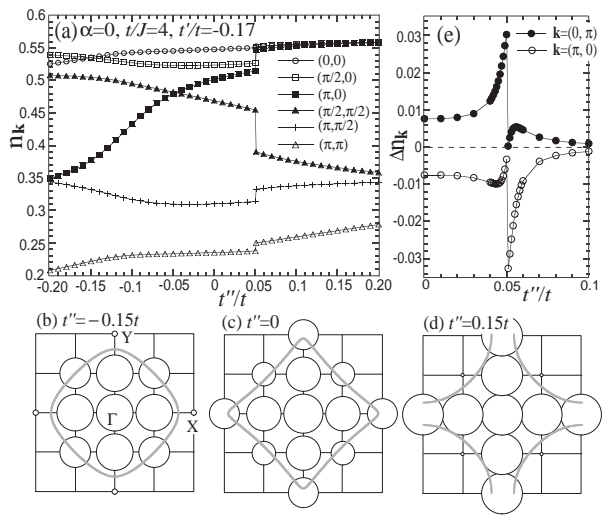}}
%
\caption{(a) $t''$ dependence of $n_\vk$ for $\alpha=0$, 
$t/J=4$, and $t'/t=-0.17$. (b)-(d) Distribution of $n_{\vk}$ 
at several choices of $t''$; the $\vk$ points with $n_{\vk}>0.35$ are 
considered and the diameter of the open circle is proportional 
to the value, $n_{\vk}-0.35$. The gray line is an inferred FS. 
(e) The change of $n_{\vk}$ by 
small $\alpha=0.01$ at $\vk=(0,\,\pi)$ and $(\pi,\,0)$.   
}
\label{nk}
\end{figure}

To understand the origin of the sharp peaks in \fig{Dhh-t2}(b), 
we calculate the momentum distribution function 
$n_{\vk}=\frac{1}{2}\bra\sum_{\sigma}{\tilde c}_{k\,\sigma}^{\dagger} 
{\tilde c}_{k\,\sigma}\ket$ for $\alpha=0$ [\fig{nk}(a)].  
The maximal value of $n_{\vk}$ is about $0.56$. 
This comes from  strong correlation effects, 
namely the constraint that double occupation is forbidden at every site, 
which leads to $n_{\vk} \leq \frac{1}{2}(1+\delta)$. 
The distribution of $n_{\vk}$ is shown in Figs.~\ref{nk}(b)-(d) for 
several choices of $t''$; the FS is inferred under the assumption of 
the Luttinger theorem\cite{luttinger60} and may represent a typical 
FS in $-0.2\leq t''/t \lesssim -0.05$ (b), 
$-0.05\lesssim  t''/t < 0.05$ (c), and 
$0.05 < t''/t \leq 0.2$ (d), respectively. 
This change of FS as a funtion of $t''$ agrees with one expected for 
noninteracting electrons with $t,t'$, and $t''$  
as emphasized before.\cite{stephan91,singh92,miyanaga94} 
Around $t''/t\approx  -0.15$ the FS is expected to cross the point 
$\vk=(\pi/2,\,\pi/2)$. But an appreciable change of $n_{\vk}$ 
is not seen in \fig{nk}(a). 
This is probably due to a finite size effect since the 
total number of electron has to be distributed over a small number of 
$\vk$ points constrained by the geometry of the cluster. 
A jump of $n_{\vk}$ is seen at $t''/t=0.05$ in \fig{nk}(a). 
This jump itself is probably a finite size effect, which 
we interpret as a level crossing between 
a state with an electron like FS [\fig{nk}(c)] and a state with 
a hole like FS [\fig{nk}(d)]. 
The free energy shows a cusp there as a function of $t''$. 
Moreover, the jump disappears for $\alpha \neq 0$ 
and $n_{\vk}$ changes smoothly, which will be due to realization 
of an ``intermediate'' state in the sense that the FS will close 
along the $k_{x}$ direction (electron like) and open along the 
$k_{y}$ direction (hole like). 
To confirm that an electron like FS actually 
changes into a hole like FS near $t''/t=0.05$ 
as inferred from Figs.~\ref{nk}(a)-(d), 
we look at the change of $n_{\vk}$ at $(\pi,\,0)$ and $(0,\,\pi)$ 
by small $\alpha$, which we define as 
$\Delta\! n_{\vk}= n_{\vk; \alpha=0.01} -  n_{\vk; \alpha=0}$. 
In \fig{nk}(e), we see 
an appreciable change of $\Delta\! n_{\vk}$ around $t''/t=0.05$  
and a positive (negative) sign of $\Delta\! n_{\vk}$ at $(0,\,\pi)$ 
[$(\pi,\,0)$]. 
This is expected when a FS is close to $(\pi,\,0)$ and 
$(0,\,\pi)$ at $t''/t=0.05$, where the small $\alpha$ moves  
the $(0,\,\pi)$ [$(\pi,\,0)$] point inside (outside) the FS. 
Therefore, considering that $\Delta\!C_{m}$ shows a sharp peak 
at $t''/t=0.05$ [\fig{Dhh-t2}(b)], we conclude that an 
electronic state with a FS being near $(\pi,\,0)$ and $(0,\,\pi)$ 
has strong correlations toward orientational symmetry breaking.

Since our anisotropy ($\alpha=0.01$)  
is  a weak perturbation to the original $t$-$J$ model, 
$\Delta\!C_{m}$ can be a linear response quantity.\cite{kubo57} 
This actually holds for most of $t''$. 
When $t''$ approaches $0.05t$, however, the linear susceptibility 
grows rapidly and the value of $\alpha=0.01$ is no longer 
in the linear response regime;  
$\Delta\!C_{m}$ begins to show saturation as a function of $\alpha$.

So far we have investigated the anisotropy of hole-hole correlation 
as a function of $t''$. 
A similar enhancement of the anisotropy 
would be expected as a function of $t'$ (with $t''=0$), 
since an electron like FS may cross $(\pi,\,0)$ and $(0,\,\pi)$, and 
evolve to a hole like FS with decreasing $t'(<0)$. 
However, $n_{\vk}$ did not indicate a hole like FS, but  
an electron like FS [\fig{nk}(c)] in a wide range 
$-0.1\gtrsim t'/t \gtrsim -0.43$. 
This is probably due to larger finite size effects at  
smaller $t'$, since $n_{\vk=(0,\,0)}$ decreases drastically 
at $t'/t \approx -0.44$ and becomes smaller than $n_{\vk}$ at 
$\vk=(\pi/2,\,0), (\pi,\,0)$, and $(\pi/2,\,\pi/2)$, 
indicating realization of a hole pocket around $\vk=(0,0)$.  
This is not expected in infinite systems for such parameters 
and in fact does not occur in the slave-boson mean-field analysis. 
Regardless of this possible finite size effect, 
$\Delta\!C_{m}$ shows a pronounced enhancement near $t'/t=-0.43$,  
where we expect a FS to be closest to $(\pi,\,0)$ and $(0,\,\pi)$ 
from the same observation as \fig{nk}(e).

The correlation of orientational symmetry breaking found in \fig{Dhh-t2}(b) 
is  identical to the $d$FSD correlation that was 
discussed in the slave-boson mean-field analysis of the 
$t$-$J$ model\cite{yamase00,yamase04b} 
and in various other models.\cite{metzner00,valenzuela01,wegner0203,metzner03,kee0304,neumayr03,yamase05a}  
A crucial point of the present study is that the $d$FSD 
correlation is found in the exact diagonalization 
where strong correlation effects are treated exactly.  
Hence the $d$FSD is relevant even to a     
strongly correlated regime and in this sense is a generic feature 
when a FS is close to $(\pi,\,0)$ and $(0,\,\pi)$.

While the present calculation has been done at a fixed hole density, 
the $d$FSD correlation may be prominent 
in a certain hole doping region. 
The slave-boson analysis of the $t$-$J$ model showed that 
the $d$FSD correlation was  noticeable in $0\leq \delta\lesssim 0.20$ 
for $t/J=4, t'/t=-1/6,t''/t=0$ and became stronger with decreasing 
$\delta$.\cite{yamase00}

The strong correlation effects are treated exactly in this 
study. But the system size is finite. Does a spontaneous $d$FSD 
take place when the system size becomes infinite?  
This is an open question. 
Since the $d$FSD competes with other 
instabilities such as $d$-wave superconductivity, there is 
a possibility that a spontaneous $d$FSD is  hindered 
by a more dominant instability.\cite{yamase00,honerkamp02}

Even if a spontaneous $d$FSD does not occur, 
appreciable correlations of the $d$FSD make a 
system sensitive to a small external anisotropy, which leads 
to a noticeable $d$FSD --- 
the FS is softened.\cite{yamase00,metzner03,yamase04b}  
This scenario was proposed for La-based cuprates 
from strong band parameter dependence 
of the $d$FSD correlation [see \fig{Dhh-t2}(b)].\cite{yamase00}  
The shape of the FS was interpreted in terms of the 
$d$FSD under the assumption of a coupling to the low-temperature 
tetragonal lattice distortion that gave a small external anisotropy 
to the electron system.\cite{yamase00} 
Magnetic excitations were investigated 
on the basis of the $d$FSD scenario and turned out to catch essential 
features of neutron scattering data.\cite{yamase010203} 
Recently anisotropic neutron scattering signals were observed   
for untwinned yttrium barium copper oxides,\cite{hinkov04} 
and effects of the $d$FSD will be interesting for this material also.

The $d$FSD correlation is the same 
as electronic ''nematic'' correlations\cite{kivelson98} 
from the view of symmetry --- orientational symmetry breaking of 
a square lattice. 
The possibility of nematic order has been discussed for high-$T_{c}$ 
cuprates.\cite{kivelson03} 
It is, however, currently envisaged as melting of spin-charge stripe order; 
the underlying physics is different from the $d$FSD. 
In the stripe scenario, some interaction with large momentum 
transfer near $\vq=(\pi,\,\pi)$  is requisite to realize 
spin-charge stripe formation. 
On the other hand, the $d$FSD comes from forward scattering 
of electrons close to the FS near 
$(\pi,\,0)$ and $(0,\,\pi)$; stripes are not necessary to get 
the nematic order. 
It should be noted that the $d$FSD scenario does not preclude realization 
of stripes, which could be driven by interactions with large 
momentum transfers in the $d$FSD state. 
In the $t$-$J$ model, however, this is unlikely. 
The hole-hole correlation is not one-dimensional even in the 
presence of the anisotropy [inset of \fig{Dhh-t2}(b)]. In addition, 
various numerical studies of the $t$-$J$ 
model reported that stripe formation 
was not favored\cite{hellberg99,sorella02} especially 
in the presence of $t'$ and $t''$,\cite{tohyama99,white99}  
that is for realistic band parameters of cuprates.

In summary, we have investigated a tendency to orientational 
symmetry breaking of the square lattice $t$-$J$ model. 
To treat strong correlation effects rigorously, we have applied the 
exact diagonalization technique to  
a $4 \times 4$ cluster with a periodic boundary condition with two holes 
at zero temperature. 
Introducing a small anisotropy to $t$ and $J$, 
we have calculated the induced anisotropy of hole-hole correlations as 
a response. 
We have found that the response is strongly enhanced for 
particular band parameters, meaning that the system has 
a strong tendency to orientational symmetry breaking of the square lattice. 
The analysis of momentum distribution function indicates that 
this correlation develops when the FS is 
close to $(\pi,\,0)$ and $(0,\,\pi)$.  
The present study shows that the idea of a $d$FSD, 
originally proposed in the slave-boson mean-field analysis of the 
$t$-$J$ model\cite{yamase00} and 
a renormalization group analysis of the Hubbard model,\cite{metzner00} 
is relevant even to a strongly correlated regime and 
is thus a generic feature of square lattice interacting 
electron systems. Possibilities of the $d$FSD are interesting 
for high-$T_{c}$ cuprates as well as other materials.\cite{grigera04}

We are grateful to Y. Hasegawa for helpful suggestions. 
H.Y. thanks L. Dell'Anna, P. Horsch, A. Katanin, G. Khaliullin, 
and W. Metzner for enlightening discussions.  


\bibliography{main.bib}

\begin{thebibliography}{35}
\bibitem[{mis()}]{miscr}
\bibinfo{note}{$\vr=(2,3)$ [$(3,1)$] is equivalent to $\vr=(2,-1)$ [$(-1,1)$]
  because of the periodic boundary condition.}
\end{thebibliography}

\end{document}